\documentclass[9pt,shortpaper,twoside,web]{ieeecolor}
\usepackage{enumerate}
\usepackage{subfig}
\usepackage{algorithmic}
\usepackage{algorithm}
\usepackage{generic}
\usepackage{cite}
\usepackage{amsmath,amssymb,amsfonts}
\usepackage{graphicx}
\usepackage{textcomp}
\usepackage{csquotes}
\begin{document}

\title{Noise-suppressing channel allocation in dynamic DWDM-QKD networks using LightGBM}
\author{Jianing Niu, Yongmei Sun, Yongrui Zhang and Yuefeng Ji
\thanks{This work is supported by National Natural Science Foundation of China Project (Grants 61831003 and 61871051). }
\thanks{The authors are with State Key Laboratory of Information Photonics and Optical Communications, Beijing University of Posts and Telecommunications, Beijing, China.}
\thanks{The corresponding author is Yongmei Sun (ymsun@bupt.edu.cn).}}

\maketitle

\begin{abstract}
Integrating quantum key distribution (QKD) with existing optical networks is highly desired to reduce the deployment costs and achieve efficient resource utilization, and some point-to-point transmitting experiments have verified its feasibility. Nevertheless, there are still many problems in the realistic  
scenario where QKD coexists with dynamic data traffics. On the one hand, the conventional static channel allocation schemes cannot guarantee the quality of quantum channels in the presence of the time-varying noises. On the other hand, considering the complex noise generation caused by dynamic classical data traffics with variable characters, it is challenging to achieve online high-performance quantum channel assignments. To address these problems, we propose a machine learning based noise-suppressing channel allocation (ML-NSCA) scheme. In this scheme, the LightGBM based ML framework is trained to predict the optimal channel allocations with lowest noise impacts, according to which, the quantum channels are periodically reallocated to guarantee high secure key rate. To improve the accuracy and scalability of the ML framework, we also optimize the method of feature extraction during the training process. The performance evaluation results indicate that the proposed scheme can effectively resist the dynamic noise impacts in the realistic optical networks and obtain higher secure key rate with less operation complexity than the previous schemes.
\end{abstract}


\section{Introduction}
The recent advances in the optical communication have enabled many emerging applications, such as Internet of things (IoT) and smart cities\cite{ji2018towards}, but the security of them is under threat since the concept of quantum computing came up. Currently, the traditional asymmetric encryption has been demonstrated to be insecure facing quantum algorithms\cite{shor1994}. Some forms of symmetric encryption are proven to be more resistant to quantum attacks, such as “one-time pad” (OTP) and advanced encryption standard (AES) encryption\cite{alleaume2014}, but the vulnerability of secure key distribution severely restricts their applications. Luckily, the emergence of quantum key distribution (QKD) provides a feasible solution to this problem. QKD can achieve information-theoretically secure key establishment between two remote parties, and the laws of quantum mechanics guarantee that it is impossible to eavesdrop information without being discovered\cite{wootters1982, Lo1999,lo2014}. Consequently, combining QKD with symmetric encryption is a promising way to resist future advances in quantum computing. In these decades, the theoretical and practical security of QKD has been constantly improved\cite{Ma2005, Yin2016, Yin2018}, and the many experimental demonstrations also show significant breakthroughs in the secure key generation rate and the transmission distance\cite{Yuan2018, Boaron2018, Lucamarini2018, wangshuang2019}. These signs of progress have paved the way for large-scale implementations of QKD networks\cite{zhang2018large}, 
but another obstacle for widespread use of QKD is the high costs of deploying dedicated fibers. 

A promising solution of reducing the deployment costs is integrating QKD with the existing optical networks. However, it is very challenging for single-photon quantum signals to share the same fiber with the intensive classical signals. Recently, the point-to-point joint transmission of quantum signals and classical signals been researched preliminarily. One of the feasible multiplexing schemes is placing the quantum signals in the O-band, which is relatively far away from the C-band classical signals, to reduce the impacts on QKD systems\cite{Townsend1997,Nweke2005, Wang2017}, and through this scheme, the multiplexing of quantum signals with 3.6 THz classical signals over 66 km backbone fiber has been achieved\cite{Mao2018}. Whereas, the low transmission loss of the C-band has also attracted more and more interests to place both quantum signals and classical signals in the C-band (i.e., DWDM-QKD scheme)\cite{Peters2009, Eraerds2010, Patel2014, wang2015experimental, Dynes2016, Frohlich2017}. In the DWDM-QKD scheme, the impairments on QKD are more serious due to the narrow frequency spacing. Some previous researches have already proven that appropriate management of channel allocations is critical for noise suppression in DWMD-QKD systems\cite{Silva2014}, and some static channel allocation schemes have been proposed to reduce the dominant in-band impairment sources such as four-wave mixing (FWM) noise and Raman scattering noise\cite{Bahrani2018, Sun2016, Niu2018, Sun2019}. 

The existing static channel allocation schemes mentioned above can provide low-noise multiplexing plans under given numbers of signals by exhaustive searching. However, in order to enable QKD in the realistic optical networks, quantum signals are required to coexist with dynamic classical data requests. In this scenario, the fixed channel allocations cannot handle the time-varying noise interferences on QKD systems. Furthermore, considering dynamic classical data traffics with variable allocations, signal powers, service holding time, etc., the traditional way of computing low-noise channel allocations every time classical channels change is of high burden and hard to fulfill the real-time requirement in the practical implementations. Therefore, more efficient online channel allocation schemes to maintain high performance of QKD are required in the dynamic DWDM-QKD networks.

Inspired by the recent researches of utilizing machine learning (ML) to predict the performances of the unestablished paths in conventional optical networks\cite{Samadi2017, Rottondi2018, Yao2018}, Y. Ou \emph{et al}. proposed an ML-based QKD performance predicting scheme\cite{Ou2018} in 2018. In their scheme, channel reallocating is invoked if the performance of QKD is worse than the requirement. This scheme improves the robustness of QKD, but it has a limited effect on improving the secure key rate (SKR). In the short term, the quantum key pool (QKP) based key management is regarded as the most feasible configuration to alleviate the limited secure key generation rate\cite{Maeda2009, Cao2017}, in which the keys are constantly generated and stored in the QKP for later use. This configuration determines that large amounts of key generations during a period is more desirable than the robustness. Therefore, the performance predicting scheme in\cite{Ou2018} fails to solve the main concern in current QKD configurations. Besides, the real-time performance evaluation and the frequent searching of backup plans are likely to overburden the network management. As far as we know, the adaptive channel allocation scheme targeted to maximize the SKR in dynamic DWDM-QKD networks is not available so far.

In this paper, considering the physical layer impairments in dynamic DWDM-QKD networks, we propose an ML-based noise-suppressing channel allocation (ML-NSCA) scheme, and the major contributions can be summarized as follows:
\begin{enumerate}[1).]
\item  To maintain good performance under dynamic noise impairments, and reduce the computation and power consumptions, quantum channels are periodically reallocated in our scheme, and the optimal channel allocations in the next period are predicted by a LightGBM based ML framework online.
\item Unavailable information of the random data traffics in the next period causes non-deterministic noise evaluations. To address this problem, we implement the Monte Carlo simulations to make the ML learn the statistical-based optimal channel allocating strategy.
\item We propose an optimized feature extraction method to reflect the status of networks, and it is verified to be effective in improving the accuracy and scalability of the ML framework.
\end {enumerate}

\section{Problem of channel allocation in the dynamic DWDM-QKD network}
\subsection{Physical-layer constrains in typical DWDM-QKD networks}

\begin{figure}[h!]
\centering
\subfloat[]{
\includegraphics[width=0.7\linewidth]{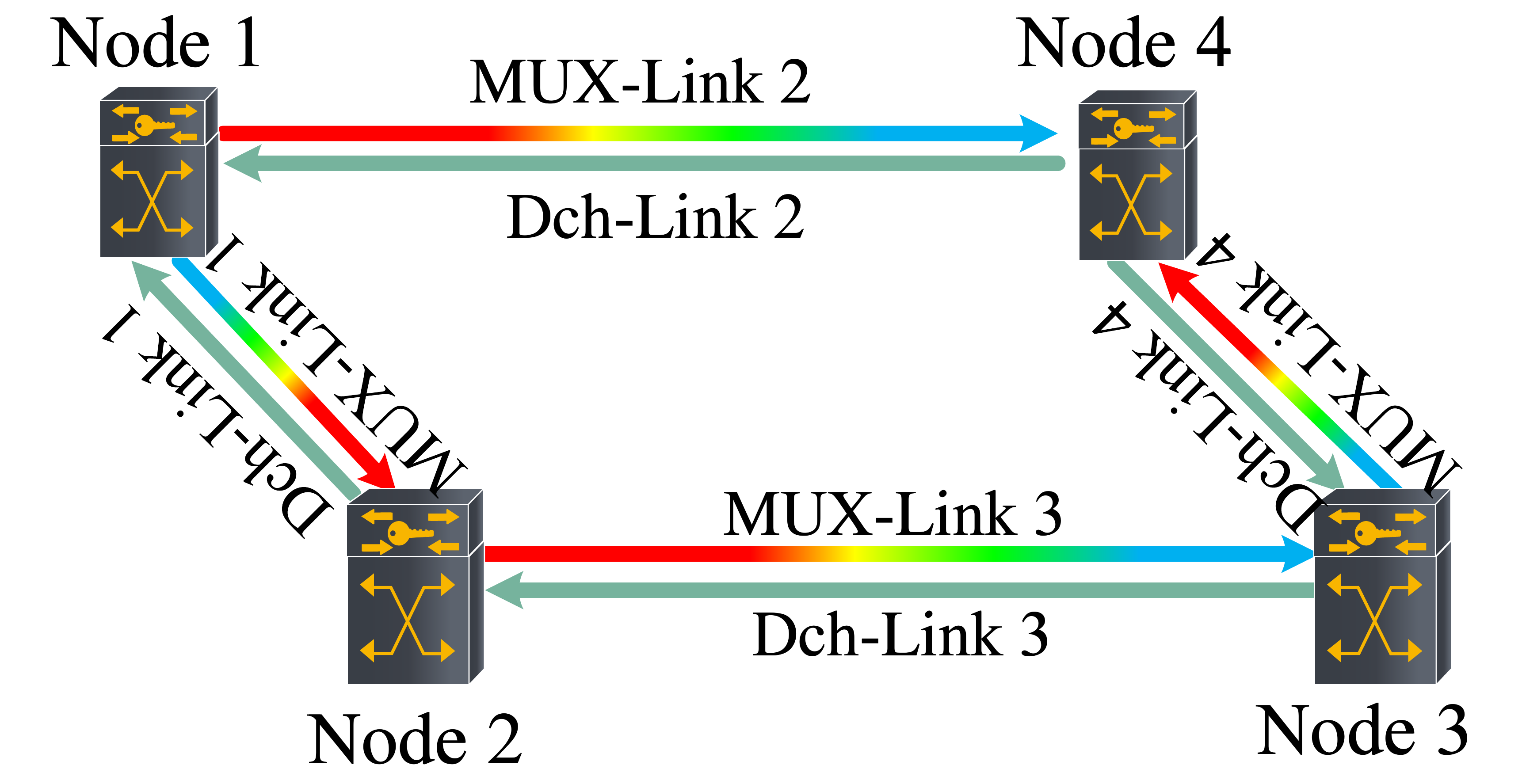}
    \label{1a}}\hfill
	  \subfloat[]{
        \includegraphics[width=0.7\linewidth]{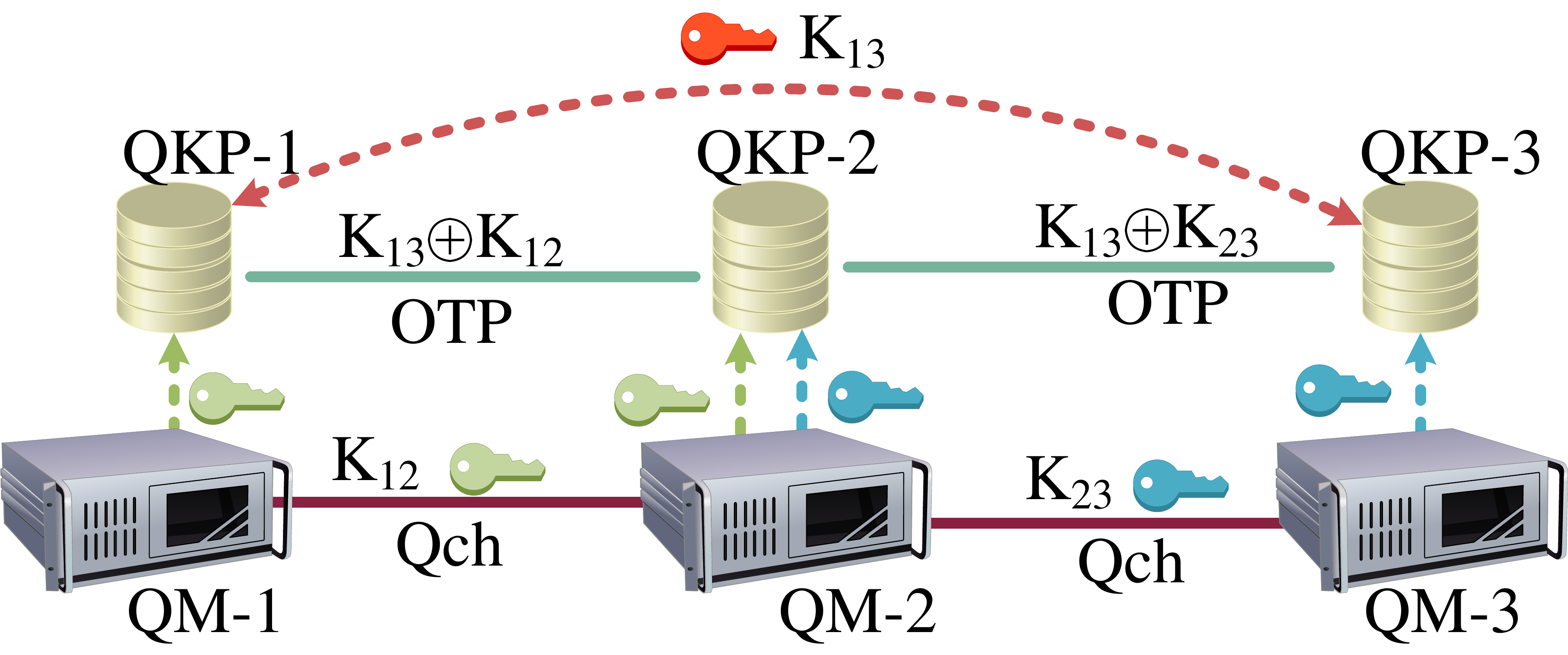}
    \label{1b}}\\
	  \caption{(a) Configuration of the DWDM-QKD network; MUX-Link: the link in which quantum signals and classical signals are multiplexed; Dch-Link: the link that only transmits the classical data signals. (b) Procedure of the trusted-relay-based key sharing, QM: quantum module.}
	  \label{figure1} 
\end{figure}

Firstly, we consider the typical construction of DWDM-QKD network. As shown in Fig. \ref{figure1}(a), the forward and backward classical communications are carried by two fibers, and the unidirectional quantum signals are transmitted through one of the fibers together with the classical signals in the same direction to avoid the harmful backward Raman scattering noise\cite{Choi2010}. To enable key sharings between any two parties in a large-scale network and avoid the extra insert loss of the optical routing node, the trusted relay approach is generally employed\cite{Qiu2014,Geihs2019}, and the process is illustrated in Fig. \ref{figure1}(b). In detail, the quantum keys are directly established between neighbor nodes and stored in the QKP, and for nonadjacent nodes, the secure keys are encrypted through OTP and delivered hop-by-hop from the source to the destination. Based on these realistic configurations in DWDM-QKD network, the following factors should be considered in the channel allocation scheme:
\begin{enumerate}[1).]
\item  Quantum channels (Qch) are unidirectional and should be multiplexed with the classical data channels (Dch) in the same direction.
\item  
The Qch allocation in each link is irrelative and independent as the key establishment between nonadjacent nodes are based on trusted relays and through Dch.
\end {enumerate}
\subsection{Noise impairments in the dynamic DWDM-QKD networks and current solutions}
In the DWDM-QKD networks, quantum signals can be impacted by several kinds of noises, including channel crosstalk noise, spontaneous Raman scattering noise, as well as FWM noise, and the deployment of erbium-doped fiber amplifier (EDFA) also causes amplified spontaneous emission (ASE) noise. Among these noises, the ASE noise can be easily eliminated by a deep notch filter or the bypass technique\cite{Aleksic2014}, and high-isolation multiplexers or narrowband filters can provide adequate isolation to suppress the channel crosstalk noise\cite{Silva2014, Niu2018}. The spontaneous Raman scattering and the FWM are two kinds of non-linear effects, and the generated noise photons may fall into the quantum channels and cannot be removed by filtering, which makes them the dominant impairment sources\cite{Peters2009}. Since the noise generation is relative to the channel frequency of classical signals and quantum signals, appropriate wavelength assignment can effectively reduce the impacts of these non-linear noises on the quantum signals.

\begin{figure}[h!]
\centering
\subfloat[]{
\includegraphics[width=0.4\linewidth]{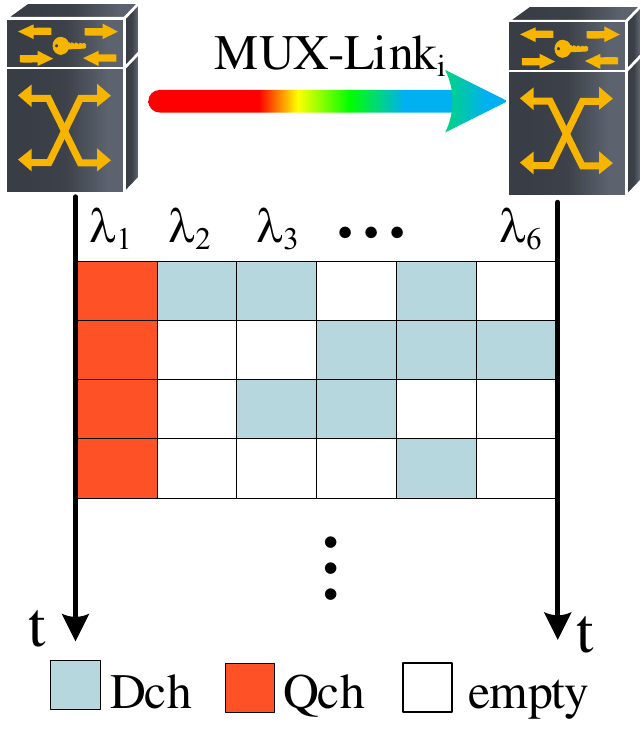}}\hfill
	  \subfloat[]{
        \includegraphics[width=0.7\linewidth]{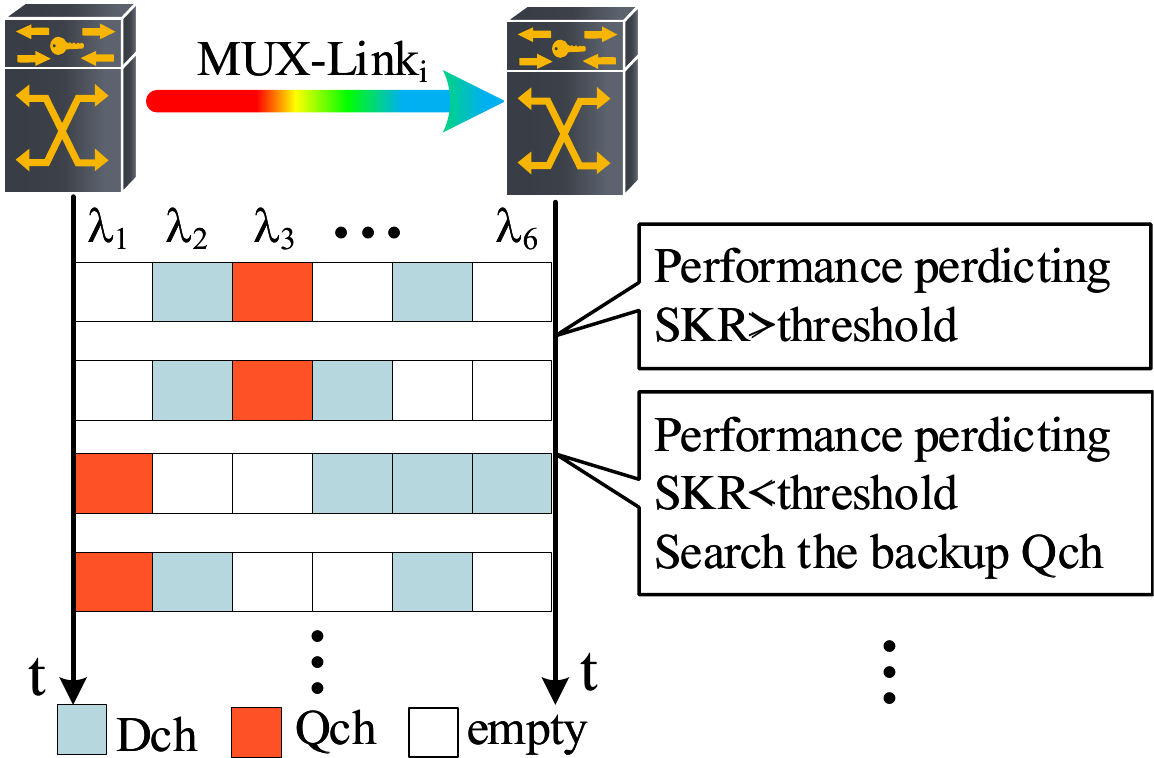}}\\
	  \caption{Illustration of the previous channel allocation schemes in the dyanmic DWDM-QKD network. (a) fixed-band scheme; (b) performance predicting scheme.}
	  \label{figure2} 
\end{figure}

In the dynamic DWDM-QKD networks, the noises are varying with the classical data traffics. Considering the complex network configurations, it is chanllenging to search the optimal channel allocations in real-time. Therefore, most of the current DWDM-QKD networks still utilize the static fixed-band (FB) channel allocation scheme, in which the Dchs and the Qchs are assigned to two separate bands (as shown in Fig. \ref{figure2}(a))\cite{Cao2018, Mao2018}, and the Qchs are fixed in the lowest wavelength to reduce the Stokes components of the Raman scattering noise\cite{Wang2017}. Obviously, the FB scheme cannot guarantee the quality of Qch in the dynamic networks. For the recently proposed performance predicting (PP) scheme shown in Fig. \ref{figure2}(b)\cite{Ou2018}, the performance of Qch is evaluated in each timeslot, and channels are reallocated if the performance is worse than the requirement. This scheme is more adaptive to the dynamic scenario, but the improvement of SKR is limited, and the real-time performance evaluation is very burdening for network management.
\section{ML-based noise-suppressing channel allocation (ML-NSCA) scheme}
To address the problem of the noise impairments on quantum signals in the dynamic DWDM-QKD network, we introduce a novel ML-based noise-suppressing channel allocation (ML-NSCA) scheme in this section. The core of the scheme is the LightGBM based ML framework, which is designed to predict the optimal channel allocations without knowing the information of future data traffics. In the training process, the dataset generation and feature extraction are optimized, so that the ML framework performs in a more accurate and scalable way.  
\subsection{Procedure of the ML-NSCA scheme}
\begin{figure}[h!]
\centering\includegraphics[width=0.7\linewidth]{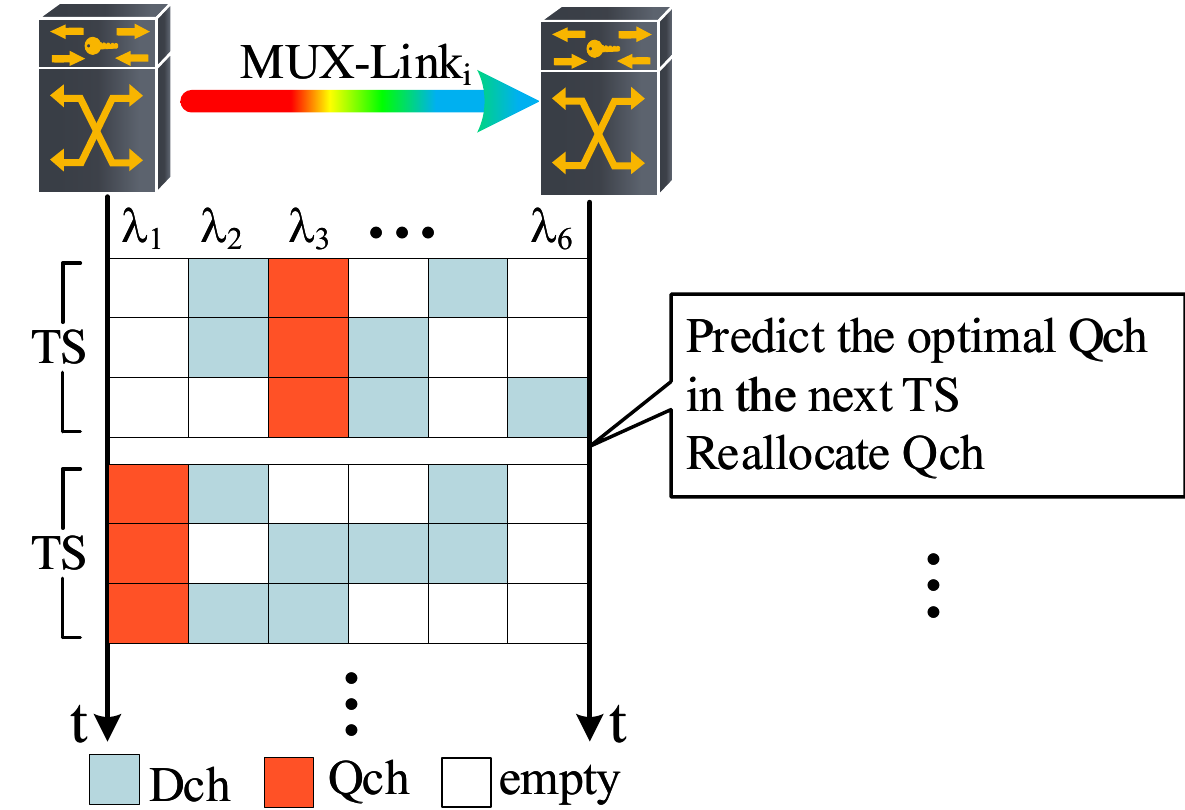}
\caption{Illustration of the ML-NSCA scheme.}
 \label{figure3} 
\end{figure}

In the ML-NSCA scheme, the Qch is reallocated periodically, as illustrated in Fig. \ref{figure3}, and the detailed procedure of the ML-NSCA scheme is described in Algorithm 1. For every timeslot, the connections of classical data requests are established first, and the allocations of Qchs remain unchanged in a fixed time window, which is assumed to be TS. After each TS, the reallocation of Qch proceeds in each MUX-link in sequence independently. According to the current state of the network, the features of each link are extracted as the input of the ML framework. Then, the trained ML framework is called to predict the optimal allocation strategy of Qch with the highest SKR in the next time window.
\begin{algorithm}[!h]
\caption{Procedure of the ML-NSCA scheme}
\begin{algorithmic}[1]
\FOR {each timeslot $t_i$}
\STATE Establish connections for data requests in $t_i$
\IF{${t_i} \equiv 0$ (mod TS)}
\FOR {each MUX-link in the network}
\STATE Release the current occupied Qch
\STATE Acquire the available channels
\STATE Extract features according to the network state for ML-based predicting
\STATE Call the trained ML framework and predict the optimal Qch reallocation scheme
\ENDFOR
\STATE Reallocate Qchs
\ENDIF
\ENDFOR
\label{al1}
\end{algorithmic}
\end{algorithm}

\subsection{Predicting the optimal channel allocation with ML}

\begin{figure*}[!h]
\centering
\includegraphics[width=0.8\linewidth]{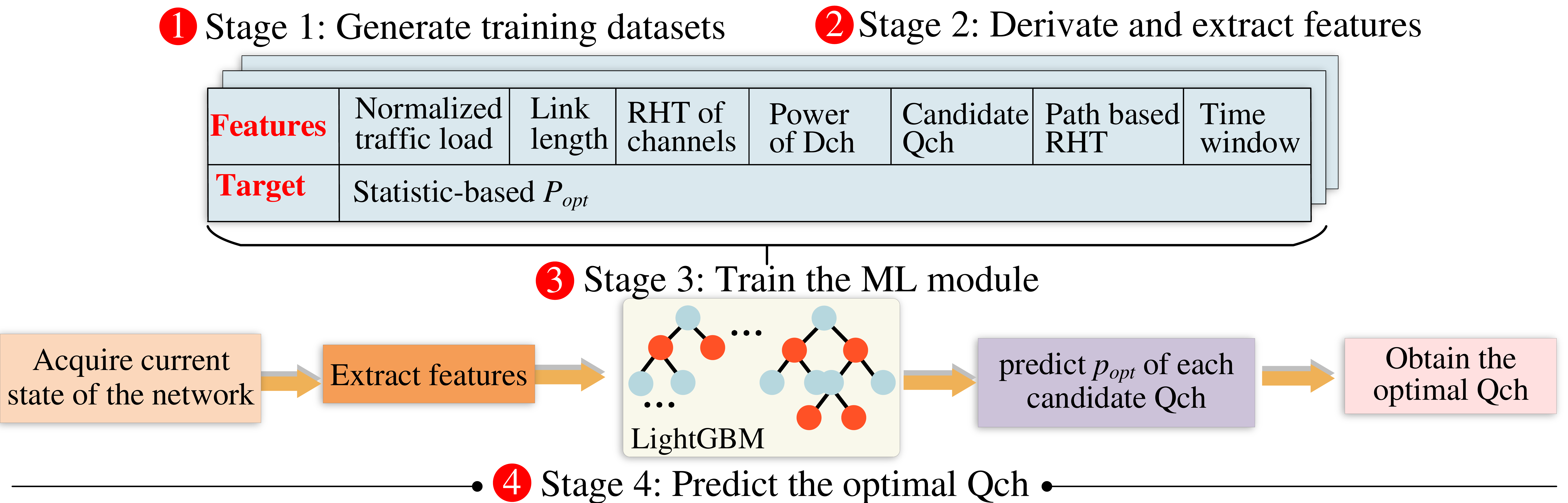}
 \caption{Procedure of the ML-based prediction of optimal Qch.}
 \label{figure4} 
\end{figure*}

In this paper, the ML module is based on the LightGBM, which is an advanced Gradient Boosting Decision Tree (GDBT) framework, proposed by Microsoft in 2017\cite{Ke2017}. The LightGBM provides better performance when the dimension of input features is relatively small, and it is proven to have less hardware requirement and high training speed. Therefore, it is suitable for our scheme, which has a few dozens of feature variables. As depicted in Fig. \ref{figure4}, predicting optimal channel allocation is carried out in four stages, which will be further described below.
\subsubsection{Generate training datasets}
In our scheme, the synthetic data are generated by simulating the data requests provisioning and theoretically evaluating the SKR of the QKD system. Different from the previous ML-based PP scheme, in which the noise estimation is done after the data requests arrive, our scheme needs to evaluate the noise impairments prior to a fixed period. Thus, the randomness of the data requests causes non-deterministic target of the training instance (i.e., the optimal Qch), which means one single simulation result is not generalizable to represent the optimal Qch in the next period. To address this problem, the Monte Carlo simulations are implemented to obtain the statistical-based training datasets. The procedure of generating data is described in Algorithm 2.
\begin{algorithm}[!h]
\caption{Procedure of the data generation}
\begin{algorithmic}[1]
\FOR {each reallocation-required timeslot $t_i$}
\STATE Release the current occupied Qch
\FOR{each MUX link in the network}
\STATE{Acquire the current state of network and the available wavelengths}
\STATE{Extract features according to the network state and construct multi-tuple $S$}
\STATE{Generate 200 sets of classical data requests for Monte Carlo simulation}
\FOR{each set of classical data requests $R$}
\FOR{each available wavelength $w_i$}
\STATE Allocate Qch at $w_i$
\STATE Establish connection for $R$ from $t_i+1$ to $t_i+TS$
\STATE Evaluate the average SKR
\ENDFOR
\IF{the highest SKR is obtained in the case of $w_i$}
\STATE The counts of each available wavelength $C_i=C_i+1$
\ENDIF
\ENDFOR
\STATE Calculate the statistical probability of each available wavelength outperforming others ${p_{opt}}=C_i/200$
\RETURN the training data $(S, {p_{opt}})$
\ENDFOR
\ENDFOR
\label{al2}
\end{algorithmic}
\end{algorithm}

We consider a dynamic scenario in which the classical data traffics follow the Poisson process and are served with shortest-path routing and first-fit wavelength assignment. At each reallocation-required timeslot, the Qch in each link is assigned to one of the available wavelengths and remains unchanged in a fixed time window (i.e., TS). Then, the average SKR during TS is evaluated by the Monte Carlo simulations (in line 6 -16 of Algorithm 2). We randomly generate 200 sets of the data traffics during TS and repeat the provisioning of each set of data traffics, and simultaneously, the average SKR of each available wavelength is evaluated. Based on the records of the wavelength with the highest SKR in each iteration, the statistical probability of each available wavelength outperforming others can be obtained (i.e., $ {p_{opt}}$), which is the target of our ML framework. 

The simulations of SKR are based on a QKD performance evaluation module, whose validity was verified by experiments in our previous researches\cite{Sun2016, Niu2018, Sun2019}. In this module, several major noise impairments are concerned, including crosstalk noise, FWM noise, Raman scattering noise, and dark count noise of the single-photon detector (SPD). We also consider the practical characteristics of the QKD system, such as the detector efficiency, detecting gate duration, and the imperfection of the interferometer and the single-photon laser source. On these bases, the lower bound of SKR can be calculated by GLLP function\cite{gottesman2004}.

\subsubsection{Derivate and extract features}
Before we start training the ML, one important work is to process the features. Redundant irrelative features not only increase the complexity of the ML framework but may also lead to over-fitting and reduce accuracy.

Preliminarily, the training data generation is carried out in the 4-node metro-scale network shown in Fig. \ref{figure1}(a), and each fiber contains 8 channels. Under the assumption that the holding time of each data service is available, totally 6 features are comprehensively considered, including the average traffic load (TL) of data services, the period of reallocating Qch (i.e., time window), the residual holding time (RHT) of each channel in all the links, the fiber length of the current processing link (i.e., The link in which the Qch prediction is processed.), the power of each channel in the current processing link, and the candidate Qch.  

\begin{figure}[h!]
\centering
\includegraphics[width=0.9\linewidth]{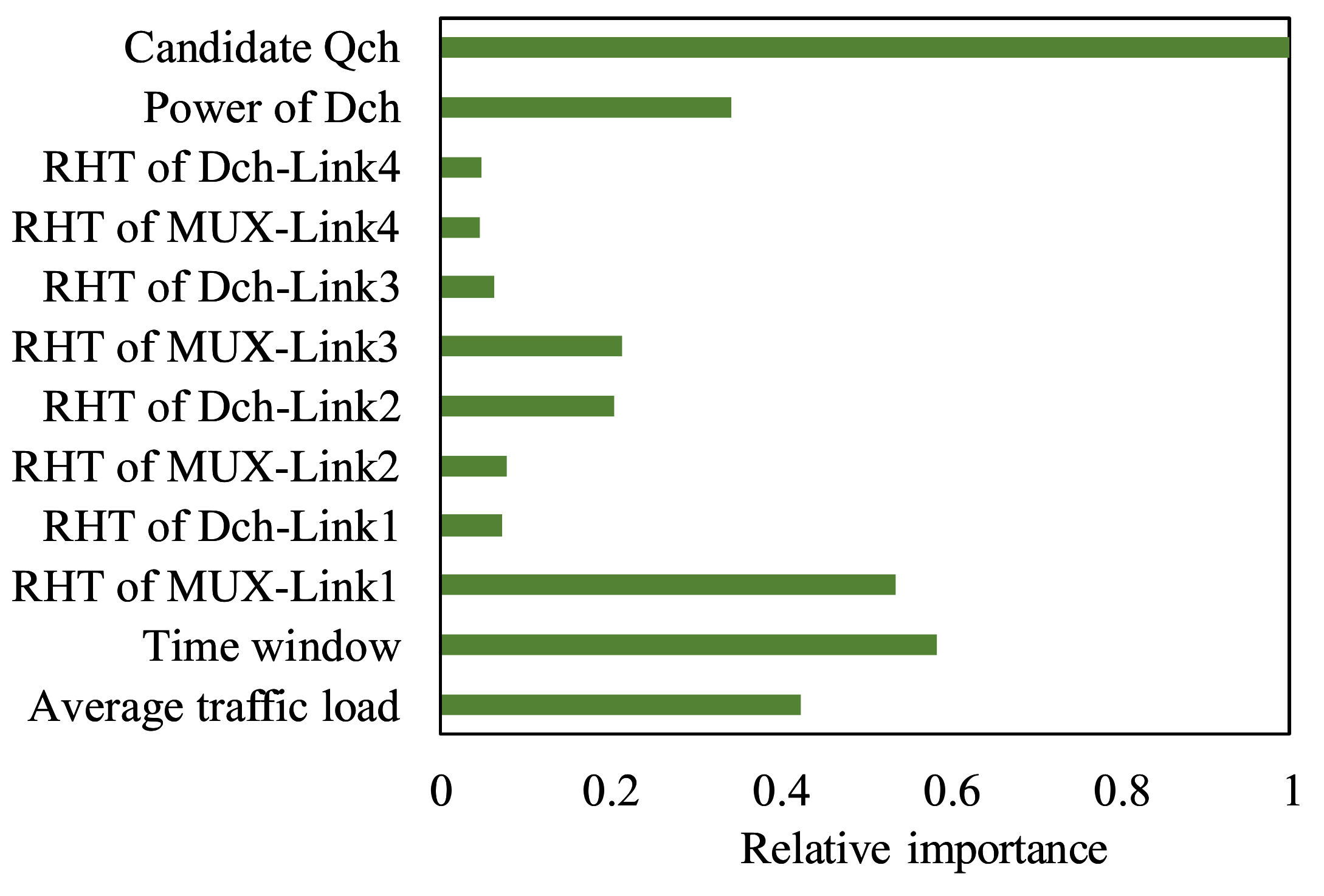}
\caption{The relative improtance of the features, and the values here are normalized by dividing the maximum value.}
 \label{figure5} 
\end{figure}

\begin{figure}[h!]
\centering
\subfloat[]{
\includegraphics[width=0.9\linewidth]{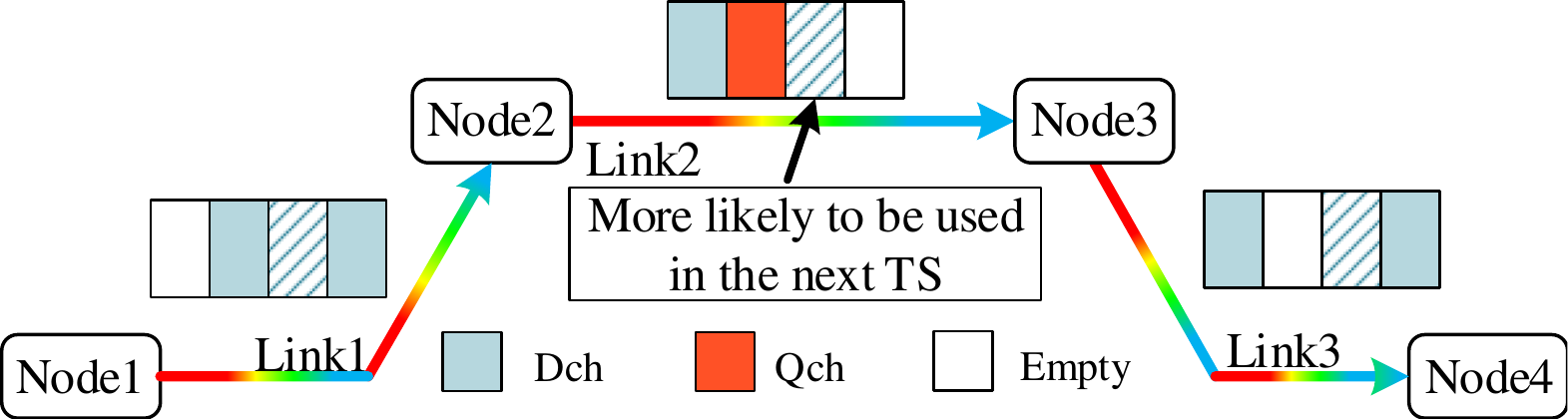}}\\
	  \subfloat[]{
        \includegraphics[width=0.9\linewidth]{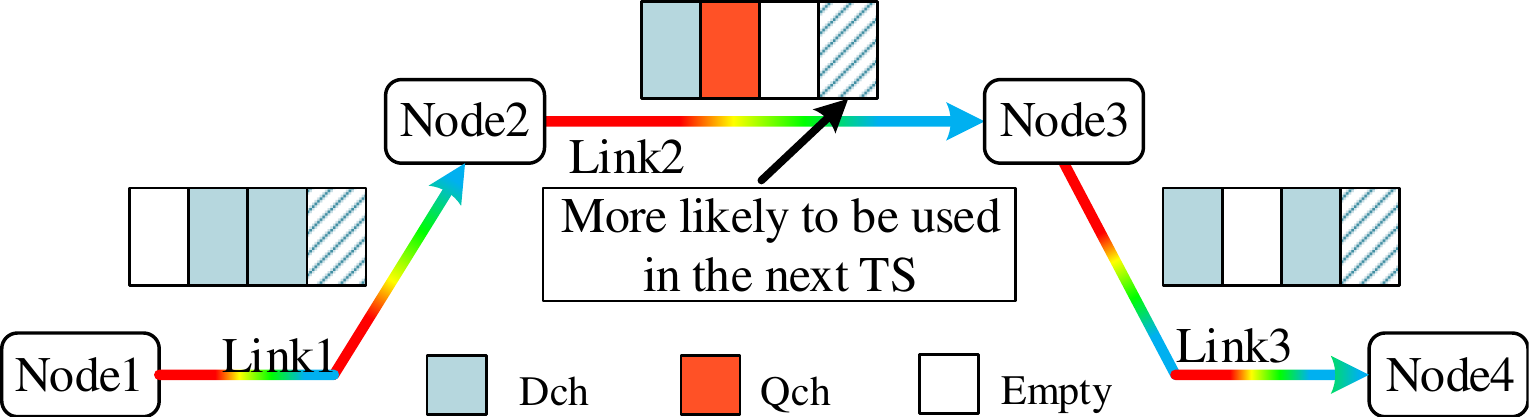}}\\
	  \caption{Illustration of the lightpath establishment of data requests in the condition of wavelength continuity constrain; (a) case 1; (b) case 2.}
	  \label{figure6} 
\end{figure}

The importance of each feature can be computed by the embedding function in the LightGBM framework. Here, we take the training datasets collected in MUX-Link1 as an example, and the results are shown in Fig. \ref{figure5} \footnote{The importance of the link length is not shown in Fig. \ref{figure5}, because it is a constant variable in this sample and does not affect the output of ML.}. We can see that the candidate Qch, time window, the RHT of the MUX-link1, average TL, and the power of Dch are predominantly related to the output of ML. Besides, the RHT of neighbor links (i.e., MUX-link3 and DC-only link2) are more relative than the RHT of other links, which is resulted from the wavelength continuity constraint of establishing the connection for classical data requests. As illustrated in Fig. \ref{figure6}, the wavelength continuity constraint requires that the lightpath occupies the same wavelength in all the links\cite{Zang2000}. Therefore, even the channel occupancies of Link2 in Fig. \ref{figure6}(a) and Fig. \ref{figure6}(b) are the same, the Qch would suffer different noise impairments in the next TS influenced by the neighbor links.

\begin{table*}[!h]
\renewcommand{\arraystretch}{1.3}
\caption{Extracted Feature Subsets}
\label{table2}
\centering
\begin{tabular}{p{100pt}p{15pt}p{70pt}p{65pt}p{65pt}}
\hline
  &\bfseries S1 &\bfseries S2&\bfseries S3&\bfseries S4\\
\hline
  Link length & \checkmark &\checkmark&\checkmark&\checkmark\\
 Average traffic load (TL) & \checkmark &\checkmark&\checkmark&normerazed \\
 Time window (TS) & \checkmark &\checkmark&\checkmark&\checkmark \\
 RHT in all links & \checkmark &RHT of the processing link& RHT of the processing link and the RHT\_path & RHT of the processing link and the RHT\_path \\ The power of each Dch in the current link & \checkmark &\checkmark&\checkmark&\checkmark \\
 Candidate Qch & \checkmark &\checkmark&\checkmark&\checkmark \\
\hline
\bfseries Size of each subset & 76 &20&28&28 \\
\hline
\end{tabular}
\end{table*}

Based on the analysis above, we extract four subsets of features, which are listed in TABLE  \ref{table2} (S1 to S4). S1 includes all the original features. S2 excludes the RHT of all the links except the current processing link. 
S3 further considers the continuity of each channel in the path passing through the current processing link. The continuity here reflects the occupancy state of the channel in each link of the path, and to evaluate it uniformly, we define a character called path-based RHT (RHT\_path), which can be expressed as 
\begin{equation}
\label{e1}
RHT\_pat{h_n} = {\mbox {Maxmum}}:~~RHT_n^m~~~~(m \in M)
\end{equation}
where  ${RHT_n^m}$ is the RHT of the link $m$ in the channel $n$, $M$ is the set of the links along the path. The channel is available to establish the lightpath only if it is not occupied in any links along the path, so the maximum RHT of all links represents the RHT of the path. Then, considering the current processing link can be involved to establish multiple paths, the RHT\_path of all the paths are averaged.    

In the subset S4, the TL is normalized to express the average TL carried by each link. This derivation is under the consideration that the utilization of each link is different according to the connections in the topology. The normalized TL (i.e., Nor\_TL) of the link $m$ can be calculated as 
\begin{equation}
\label{e2}
Nor\_T{L_m} = \frac{{{p_m}\lambda }}{\mu } = {p_m} \cdot TL
\end{equation}
where $\lambda$ is the average arriving rate of the classical data traffic, and the average holding time of each classical data service is represented by $1/\mu$. $p_m$ is the probability of the arriving data traffic passing through the link $m$. For example, in the 4-node topology in Fig. \ref{figure1}(a), there are 12 kinds of combinations of source and destination nodes, and three of them require MUX-Link1 to establish connections, so ${p_1}$ is 0.25. The performance of the four subsets of features will be evaluated in the next section, and the subsets with better accuracy and less complexity would be adopted in the ML-NSCA scheme.
\subsubsection{Train the ML framework and predict the optimal Qch}
Apart from the relevance of the features, the performance of the ML framework is also affected by several parameters, such as the training iteration, the learning rate, the number of leaves, the maximum tree depth, and so on. Among them, two particular parameters of LightGBM are the number of leaves and the maximum tree depth. The former is directly related to the complexity of the ML framework, and the latter should be larger than ${\log _2}(num\_leaves)$ to avoid over-fitting based on experiences. By adjusting the value of each parameter step by step and testing them with ten-fold cross-validation, the combination with the best tradeoff between accuracy and time-consuming will be our final choice.

Based on the trained ML framework, the $p_{opt}$ of each available wavelength is predicted according to the input features. The available wavelength with the highest $p_{opt}$ is most likely to have minimal impacts of noises during the next time window, so it is selected as the optimal Qch. The ML framework here output a performance evaluation index $p_{opt}$ rather than the optimal Qch, which provides an opportunity to design more flexible algorithms for selecting Qch.

\section{Performance evaluation of the ML-NSCA scheme}
In this section, we first analyze the effectiveness of our feature extraction method and evaluate the accuracy of the trained ML framework to predict the optimal Qch. Then, to verify the superiority of the ML-NSCA scheme, the average SKR of the QKD system is numerically assessed in different channel allocation schemes. 

\subsection{Analysis of the feature extraction method}

For each feature subset described in TABLE \ref{table2}, we test the accuracy of ML through ten-fold cross-validation based on the datasets generated in the 4-node network, and the root mean squared error (RMSE) are reported in TABLE \ref{table3}.
\begin{table}[!h]
\renewcommand{\arraystretch}{1.3}
\caption{RMSE comparision of different feature subsets}
\label{table3}
\centering
\begin{tabular}{p{30pt}p{40pt}p{40pt}p{40pt}p{40pt}}
\hline
\bfseries  & \bfseries S1& \bfseries S2& \bfseries S3& \bfseries S4\\
\hline
\bfseries RMSE  &  0.0314&0.0913& 0.0298&0.0287\\
\hline
\end{tabular}
\end{table}

For S1, it contains the largest number of elements, and the RMSE is 0.0314. S2 contains fewest elements, but the RMSE is the highest due to the leakage of critical features. The subset S3 and S4 have similar RMSE, which is lower than that of S1. These results indicate that the link information required for predicting can be well reflected by the proposed feature derivation and extraction method in S3 and S4. What’s more, the removal of some redundant features in S1 also improves the accuracy and reduces the complexity of the ML framework.
\begin{figure}[h!]
\centering
\subfloat[]{
\includegraphics[width=0.45\linewidth]{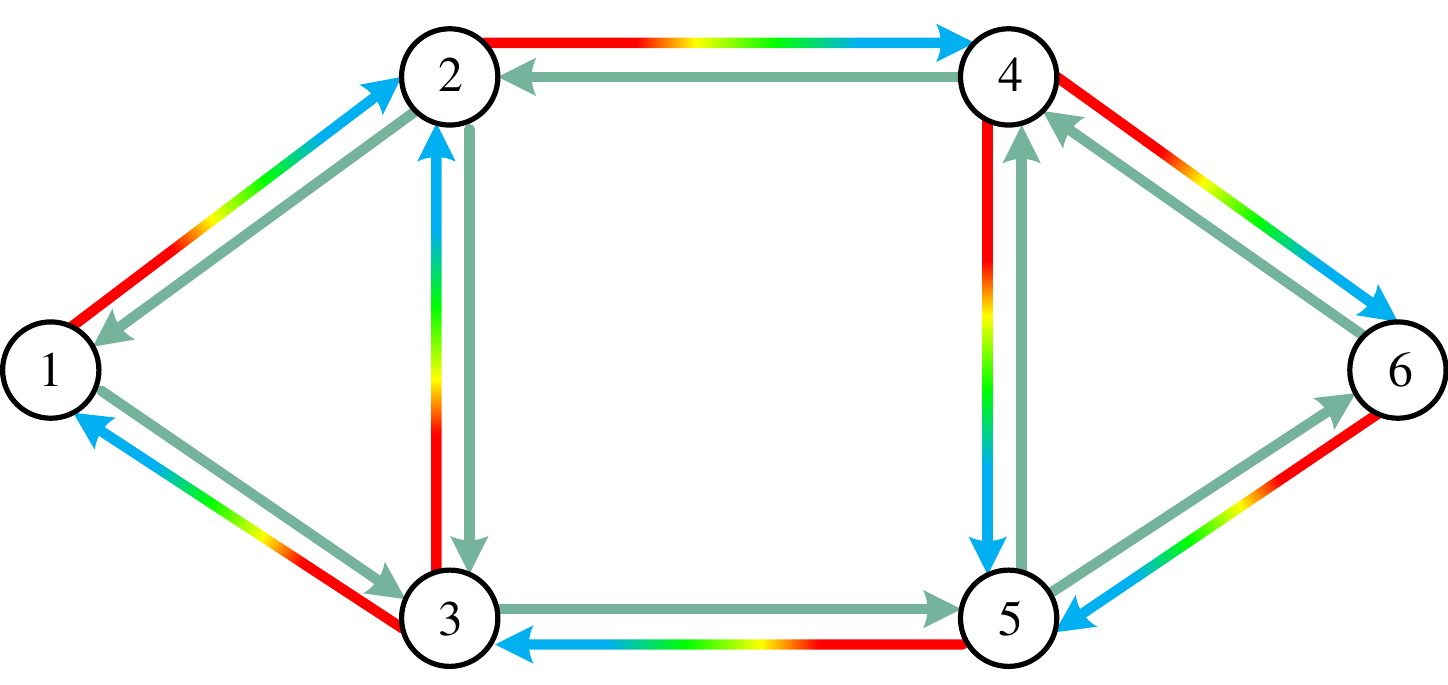}}\hfill
	  \subfloat[]{
        \includegraphics[width=0.48\linewidth]{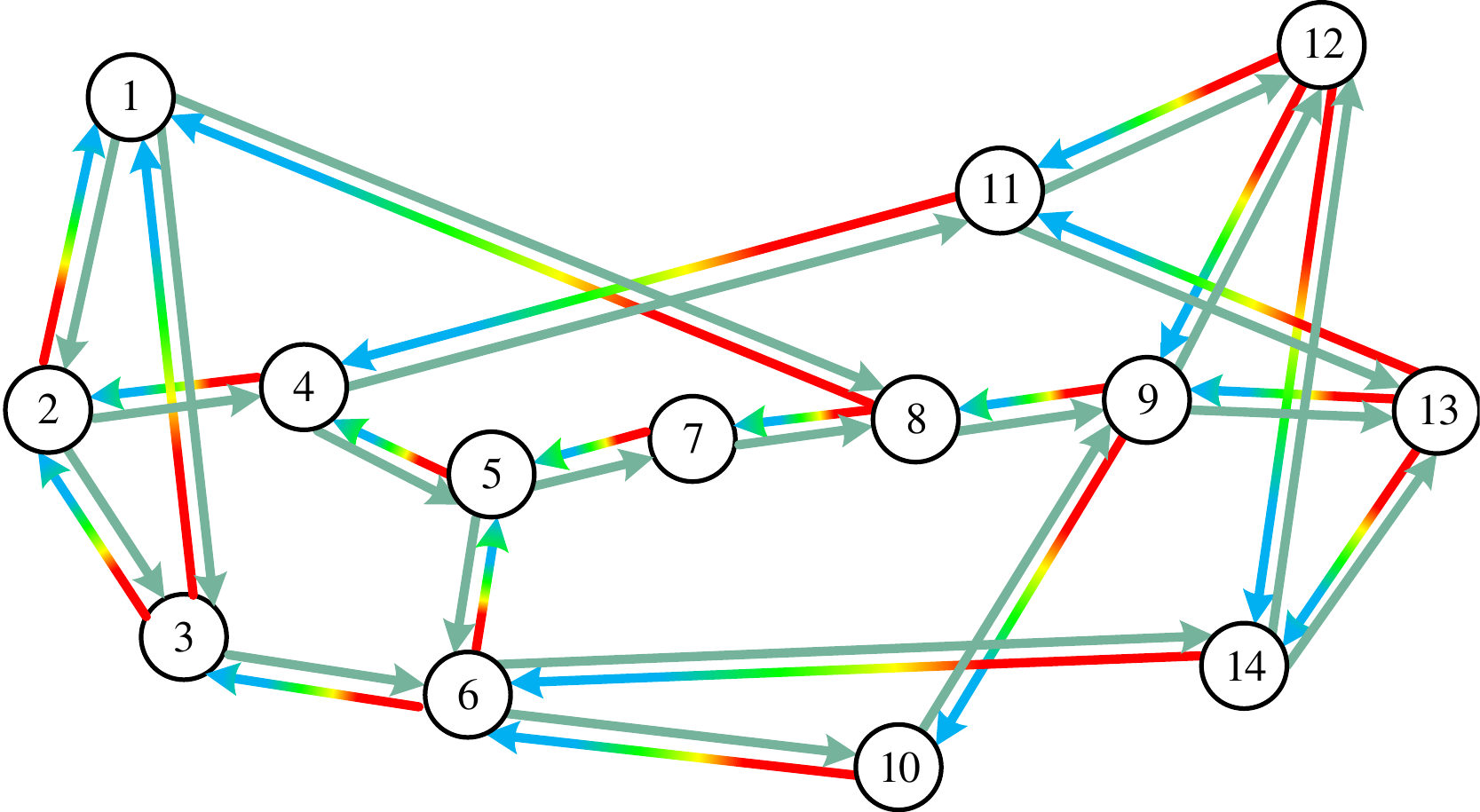}}\\
	  \caption{Network topology; (a) 6-node DWDM-QKD network; (b) 14-node DWDM-QKD network.}
	  \label{figure7} 
\end{figure}

Although S3 and S4 have similar performances, S4 can reflect the features of each link independently and is irrelevant to the network topology because it processes the TL of the network to represent the average load in each link. 
Therefore, through the feature extraction method in S4, the trained ML framework can be applied in other kinds of topologies. To evaluate this scalability, we generate two test datasets in the 6-node network and the 14-node NSFNet shown in Fig. \ref{figure7}, based on which, the ML framework trained by the datasets generated in the 4-node network are tested. The results are shown in TABLE \ref{table4}, which verify that the feature extraction method in S4 significantly improves the RMSE in different topologies.

\begin{table}[!h]
\renewcommand{\arraystretch}{1.3}
\caption{RMSE tested in the 6-node network and the 14-node network}
\label{table4}
\centering
\begin{tabular}{p{50pt}p{85pt}p{85pt}}
\hline
\bfseries  & \bfseries 6-node network& \bfseries 14-node NSFNet\\
\hline
\bfseries S1& \multicolumn{2}{p{170pt}}{Cannot be generalized to other topologies. }\\

\bfseries S2&0.168&0.189\\

\bfseries S3&0.139&0.163\\

\bfseries S4&0.038&0.048\\
\hline
\end{tabular}
\end{table}

It can be concluded that the feature derivation and extraction method can improve the accuracy of the ML framework and effectively simplify the dimension of features. Even for the complex topology, the required features are not increased, which reduces the cost of computation and storage. Besides, the link-based feature extraction also makes the trained ML framework generalized to different topologies, so it is unnecessary to be retrained when the optical nodes in the network are added or removed, which makes the ML framework has good scalability and can also be adapted to the topology-reconfigurable networks. 

\subsection{The accuracy of predicting optimal Qchs}
After carefully adjusting, the best performance of ML can be obtained when the main parameters are set as TABLE \ref{table5}, and in this case, the RMSE can reach 0.029. 

\begin{table}[!h]
\renewcommand{\arraystretch}{1.3}
\caption{Parameter settings of the ML module}
\label{table5}
\centering
\begin{tabular}{p{140pt}p{50pt}}
\hline
\bfseries Parameter & \bfseries Value\\
\hline
Size of the training dataset& $2\times 10^6$\\

Learning rate&0.1\\

Training iteration&5000\\

Number of leaves&40\\ 

Minimum data in leaf &20\\
Max\_bin&100\\
\hline
\end{tabular}
\end{table}

\begin{figure}[!h]
\centering
\includegraphics[width=0.8\linewidth]{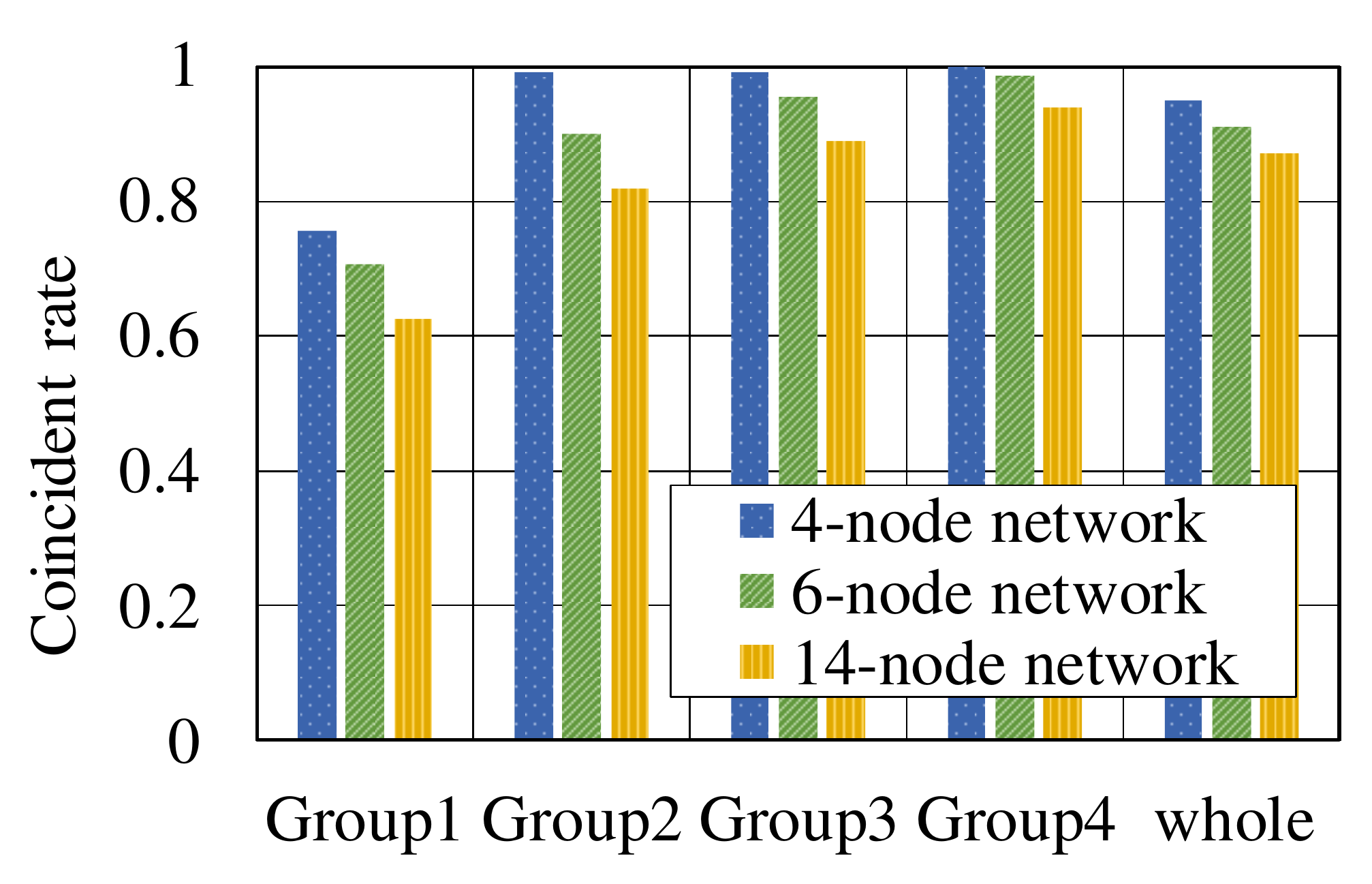}
  \caption{Coincident rate in different subsets of test data.}
  \label{figure8} 
\end{figure}

The deviation of ML directly affects the accuracy of identifying the optimal Qch. We calculate the coincident rate of the estimated optimal Qch and the true optimal Qch in the test datasets. To analyze the ability to identify the obvious instances and the ambiguous instances, we divide the whole test datasets into four groups according to the difference between the highest $p_{opt}$ and the second $p_{opt}$, which is described in TABLE \ref{table6}. The results of the coincident rate in different groups are shown in Fig. \ref{figure8}.
\begin{table}[h]
\renewcommand{\arraystretch}{1.3}
\caption{Description of dividing the test datasets.}
\label{table6}
\centering
\begin{tabular}{p{30pt} p{90pt} p{80pt}}
\hline
&\bfseries Proportion in the whole test dataset&\bfseries difference between the highest $p_{opt}$ and the second $p_{opt}$\\
\hline
\bfseries Group 1 &19\% in 4-node network; 21\% in 6-node network; 10\% in 14-node network&$<20\%$\\
\bfseries Group 2 &32\% in 4-node network; 12\% in 6-node network; 24\% in 14-node network &$[20\%, 40\%]$\\
\bfseries Group 3 &24\% in 4-node network; 19\% in 6-node network; 23\% in 14-node network &$[40\%, 60\%]$\\
\bfseries Group 4 &25\% in 4-node network; 48\% in 6-node network; 43\% in 14-node network &$>60\%$\\
\hline
\end{tabular}
\end{table}

For the 4-node network, the coincident rate in the whole test datasets can achieve 95\%. Most of the errors are generated in Group 1, in which the optimal Qch has very close $p_{opt}$ with another channel, and for Group 2, Group 3, and Group 4, the coincident rates are about 99\%. It can be concluded that the ML framework in our scheme can clearly identify the optimal Qch with outstanding performance, although some mistakes are made when it identifies the instances where the channels have similar $p_{opt}$, it is acceptable due to these channels have similar performances on SKR. Similar conclusions can also be obtained for the results in the 6-node network and the 14-node NSFNet, except for a slight decrement of accuracy when the network becomes complex. 

\subsection{Performance evaluation of improving SKR}
To verify the superiority of the ML-NSCA scheme in improving the SKR, we compare it with the FB scheme and the PP scheme, which are described in Section 2. It needs to be supplemented that for the PP scheme, the backup channel plan is not declared in \cite{Ou2018}, so the channel with lowest noise impacts is selected as the backup plan in our simulation to get the upper bound of the PP scheme. Besides, considering the fairness, the threshold of the PP scheme is set to make the times of channel reallocation equal to that in the ML-NSCA scheme. The main simulation parameters are stated in TABLE \ref{table1}.
\begin{table}[!h]
\renewcommand{\arraystretch}{1.3}
\caption{Major simulation parameters of DWDM-QKD system}
\label{table1}
\centering
\begin{tabular}{p{180pt}p{50pt}}
\hline
\bfseries Parameter & \bfseries Value\\
\hline
Frequency of the detector & 10 MHz\\
Efficiency of the SPD & 10\%\\
Dark count probability& $3\times 10^{-6}$\\
Gate duration of the SPD & 500 ps\\
Interference visibility & 95\%\\
Frequency spacing of each channel& 200 GHz\\
Bandwidth of the filter before QKD receiver & 15 GHz\\
Insert loss of the DWDM system& 8 dB\\
\hline
\end{tabular}
\end{table}

\begin{figure*}[h]
\centering
\subfloat[]{
\includegraphics[width=0.32\linewidth]{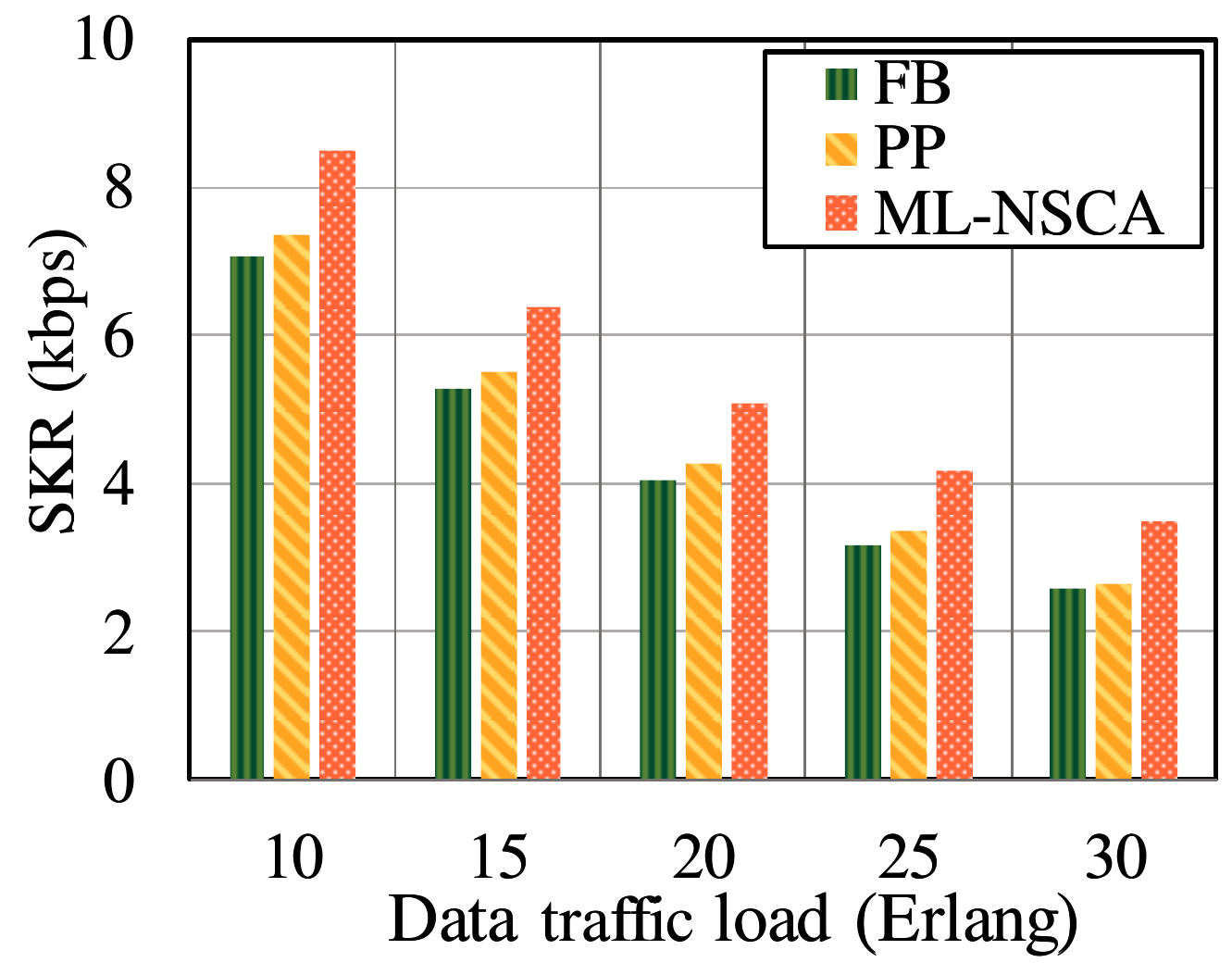}}
	  \subfloat[]{
        \includegraphics[width=0.35\linewidth]{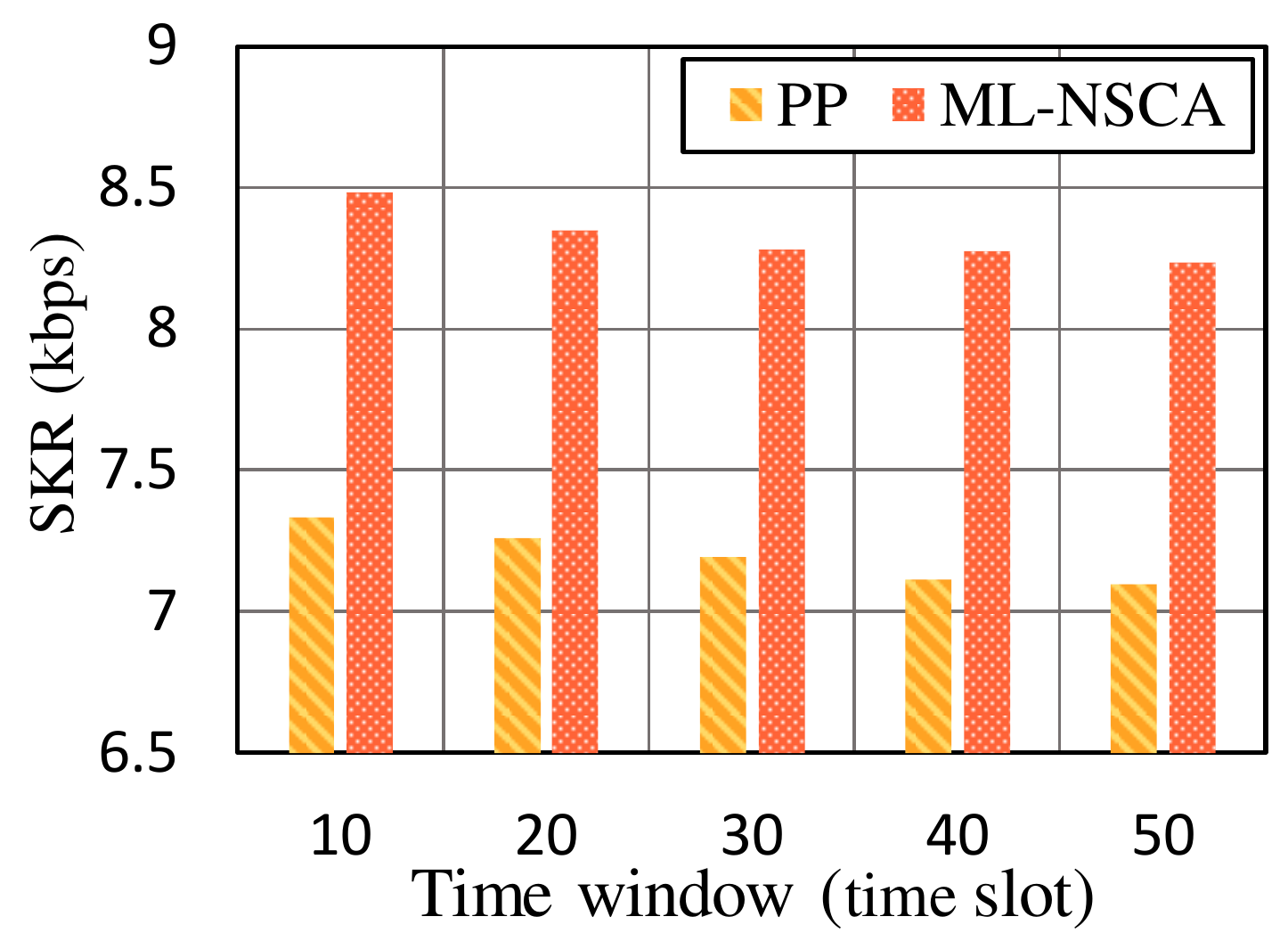}}\hfill
        \subfloat[]{
        \includegraphics[width=0.35\linewidth]{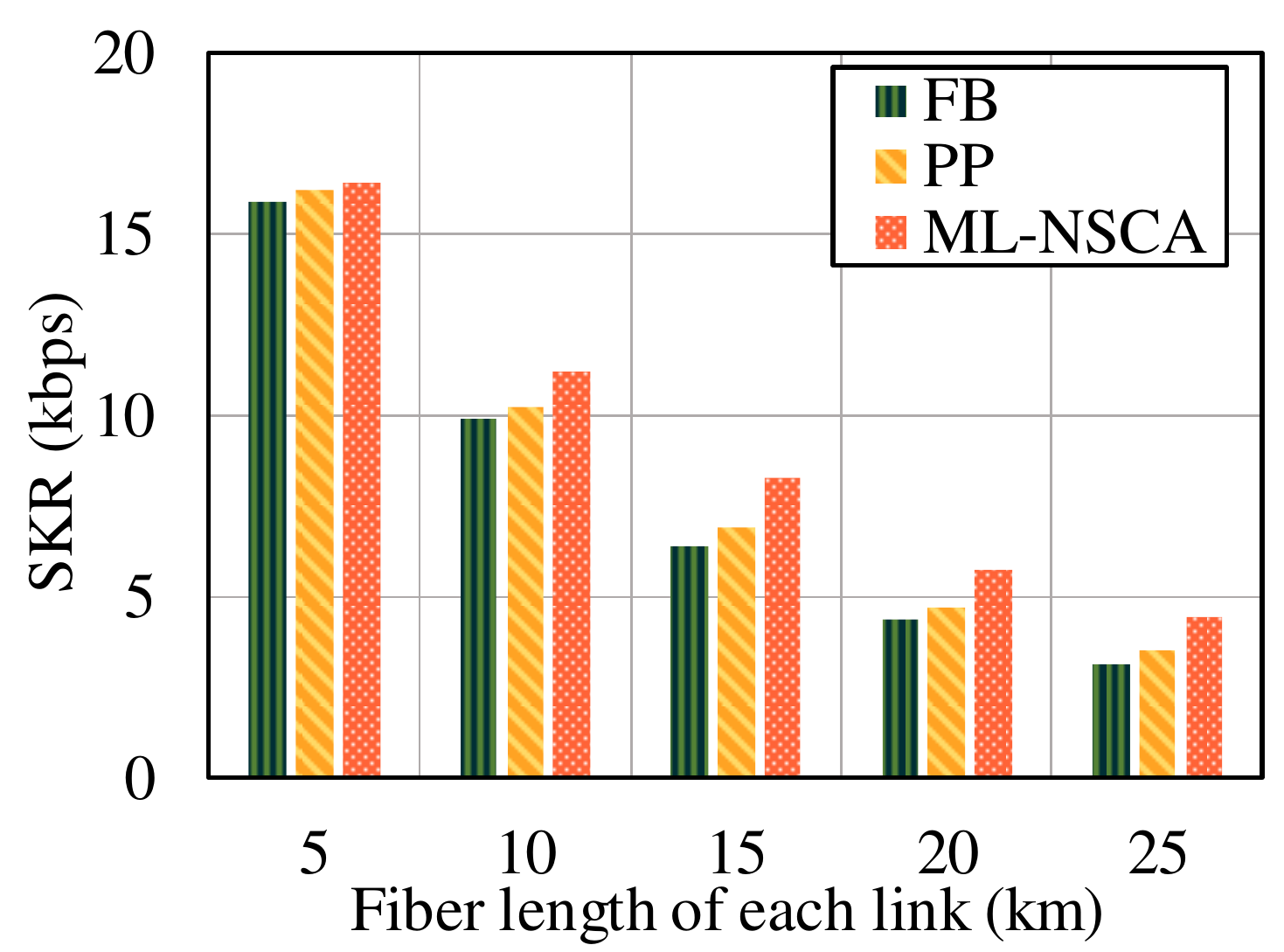}}
        \subfloat[]{
        \includegraphics[width=0.32\linewidth]{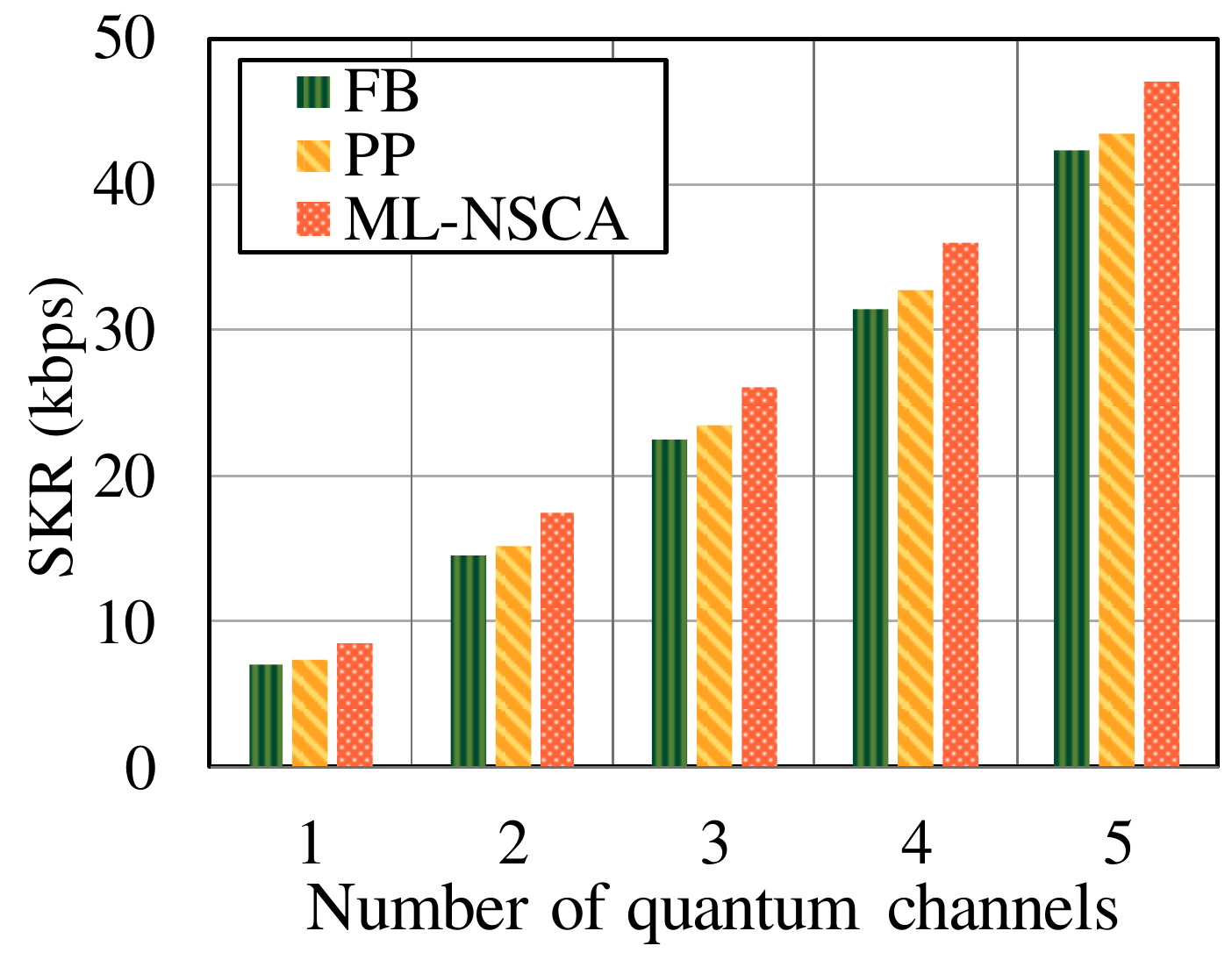}}\hfill
	  \caption{Evaluations of SKR in 4-node network; (a) SKR vs. data traffic load with $P_{Dch} =[-5, 5]$ dBm, TS=10 time slot; (b) SKR vs. configuration period with $P_{Dch} =[-5, 5]$ dBm, TL=10 Erlang; (c) SKR vs. fiber length of each link with TL=10 Erlang, TS=10 time slot, $P_{Dch} =[-5, 5]$ dBm; (d) SKR vs. number of Qchs with TL=10 Erlang, TS=10 time slot, $P_{Dch} =[-5, 5]$ dBm.}
	  \label{figure9} 
\end{figure*}

The simulation results in Fig. \ref{figure9} are obtained in the 4-node network in Fig. \ref{figure1}(a), and the four links range 5 km, 15 km, 20 km, and 30 km respectively. In each simulation, the average SKR is obtained after serving 10000 data requests, and the statistic accuracy is ensured by 100 times of repetition. 

The results of SKR versus the data traffic load are shown in Fig. \ref{figure9}(a). The SKR gradually decreases with the increment of TL, which results from more classical signals in the network generating more noises. The SKR in the FB scheme is the lowest because the quality of the fixed Qch cannot be guaranteed in the condition of time-varying noises. Both the PP scheme and ML-NSCA scheme can dynamically adjust the Qch allocations, but with the same operation complexity (i.e., the same times of channel reallocation), the PP scheme has a very limited performance of improving the SKR. Whereas the ML-NSCA scheme can obtain the highest SKR, especially, in the case of data traffic load of 30, the SKR is 35\% and 31\% higher than that in the FB scheme and the PP scheme respectively. Additionally, Fig. \ref{figure9}(b) indicates that reallocating Qch more frequently can achieve higher SKR for dynamic schemes like the PP scheme and ML-NSCA scheme, but it is also at the costs of high operation complexity. Nevertheless, our proposed scheme not only has better performance under the same operation complexity but also avoid real-time performance predicting, which lightens the burden of network management.

Fig. \ref{figure9}(c) shows the SKR versus the fiber length of each link. When the fiber length is 5 km, the noise powers generated by the classical signals are relatively small, so the improvement of SKR in the PP scheme and ML-NSCA scheme is not substantial. As the transmission distance increases, the noise interferences are more and more serious, and in this case, the proposed scheme becomes more effective. At the distance of 25 km, the SKR in the ML-NSCA scheme can achieve 4.4 kbps, which is 42\% more than the FB scheme and 26\% more than the PP scheme. In Fig. \ref{figure9}(d), we increase the number of Qchs and evaluate the total SKR in different schemes. The results of the ML-NSCA scheme are obtained by predicting first $i$ (i.e., the number of Qchs) optimal Qchs in turn. Although this method is not the global optimum solution, it can be more flexible and scalable, and the ML-NSCA scheme still obtains highest SKR under different numbers of Qchs.

\begin{figure}[!h]
\centering
\subfloat[]{
\includegraphics[width=0.65\linewidth]{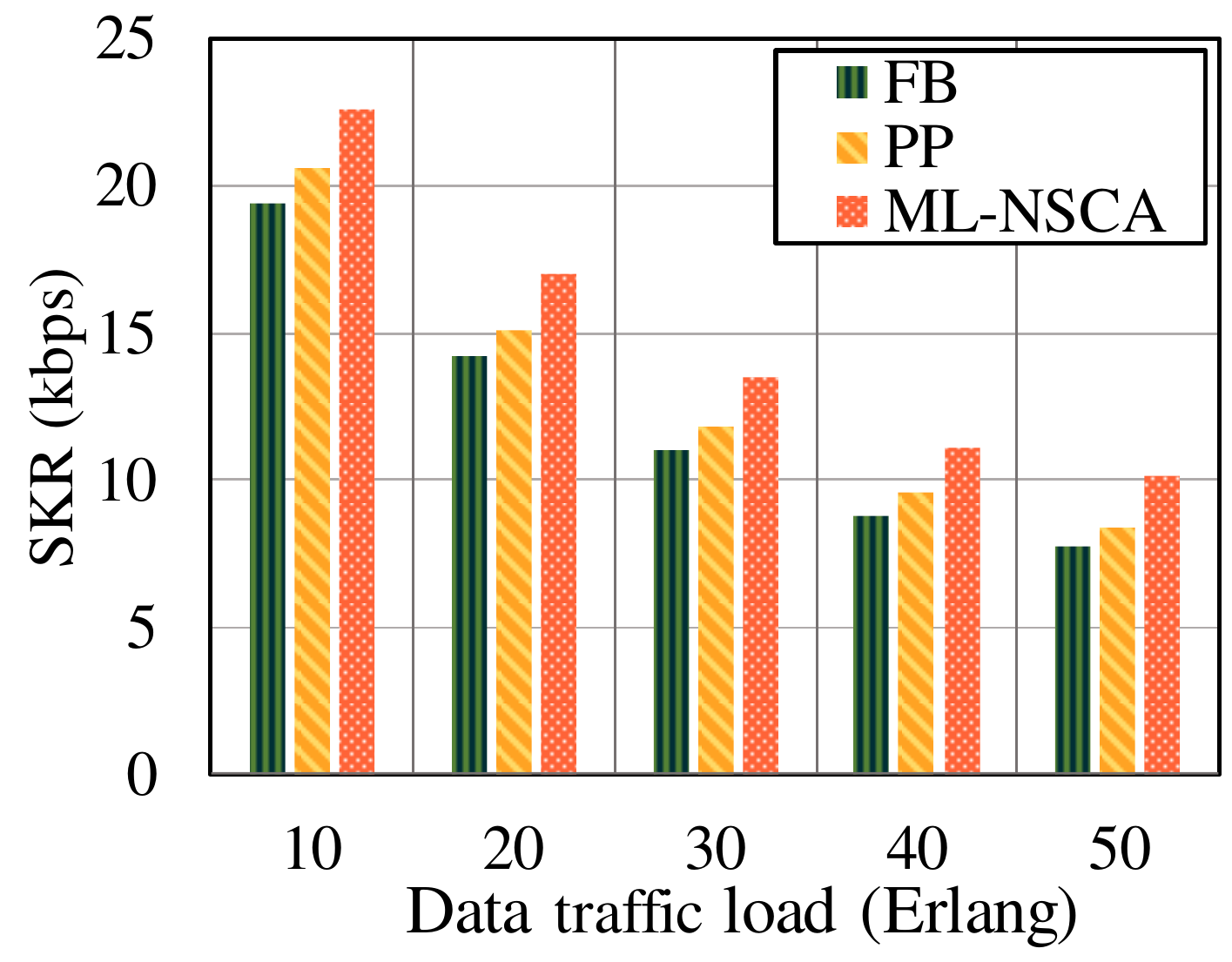}}\hfill
	  \subfloat[]{
        \includegraphics[width=0.65\linewidth]{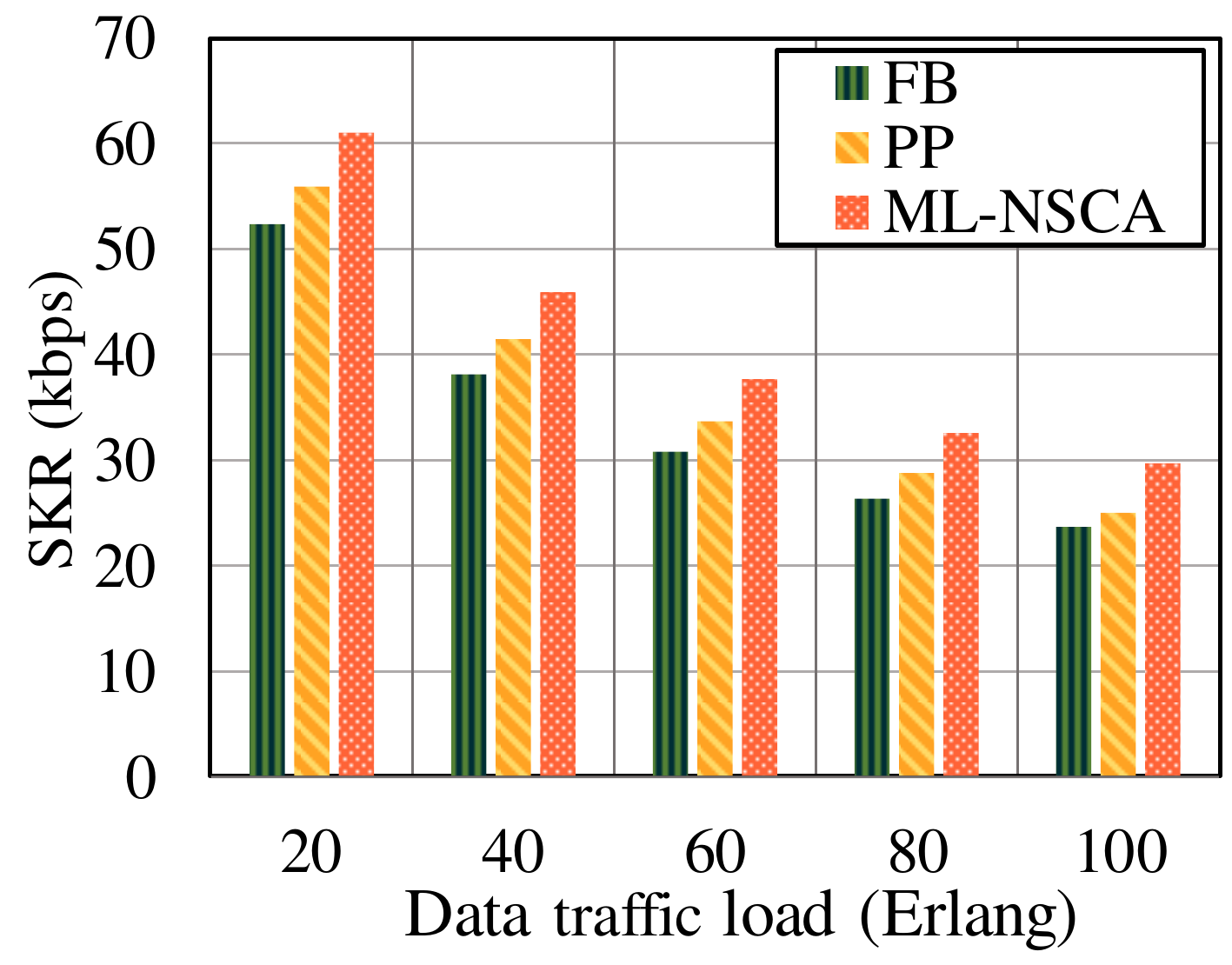}}\hfill
	  \caption{SKR vs. data traffic load in 6-node network and 14-node network; (a) SKR vs. data traffic loads in 6-node network with $P_{Dch} =[-5, 5]$ dBm, TS=10 time slot; (b) SKR vs. data traffic loads in 14-node network with $P_{Dch} =[-5, 5]$ dBm, TS=10 time slot.}
	  \label{figure10} 
\end{figure}

We also evaluate the performances of the ML-NSCA scheme in the 6-node network and the 14-node NFSNet, and the fiber length of each link is randomly set between 5 km to 30 km. The results in Fig. \ref{figure10} verify that the superiority of the ML-NSCA scheme can be maintained in different networks.

\section{Conclusion}
In this paper, an ML-NSCA scheme is proposed to reduce the noise impairments on quantum signals in the scenario of QKD being integrated into dynamic optical networks. A LightGBM based ML framework is designed to predict the optimal channel allocations with the highest probabilities to obtain better SKR in the presence of random data traffics. 
Through optimizing the feature extraction method, the ML framework can obtain high accuracy of identifying optimal channel allocations, and it also has good scalability for different network topologies.
The comparison with the existing schemes shows that the ML-NSCA scheme significantly improves the SKR than the FB scheme, and it can also effectively obtain higher SKR than PP scheme with less operational complexity. Our research here provides a feasible method to reduce the impairments on quantum signals when they coexist with dynamic data traffics, and it is meaningful for promoting the integration of QKD with the realistic optical communication networks. 
Furthermore, the feature extraction method in this paper still can be improved to obtain better performance in complex optical networks, which needs to be further researched. Besides, novel noise-suppressing channel allocation schemes with flexible recollection period will also be investigated in our future works.

\end{document}